\documentclass[a4paper,10pt]{article}
\RequirePackage[left=1in,right=1in,top=1in,bottom=1in]{geometry}
\usepackage{latexsym}
\usepackage{caption, subcaption}
\usepackage{amsmath,amssymb}
\usepackage{graphicx} 
\begin{document}
\date{}
\title{\LARGE{\bf
Noise-induced phase transition in the model of human virtual stick balancing}
}
\author{
Arkady Zgonnikov \\
National University of Ireland, Galway, University Road, Galway, Co. Galway, Ireland\\
E-mail: arkady.zgonnikov@nuigalway.ie \\
\\
Ihor Lubashevsky \\
University of Aizu, Tsuruga, Ikki-machi, Aizu-Wakamatsu City, \\ 
Fukushima 965-8560, Japan\\
E-mail: i-lubash@u-aizu.ac.jp
}

\maketitle
\thispagestyle{empty}

\begin{figure}
	\centering
	\begin{subfigure}{0.18\textwidth}
		\includegraphics[width=1.0\textwidth]{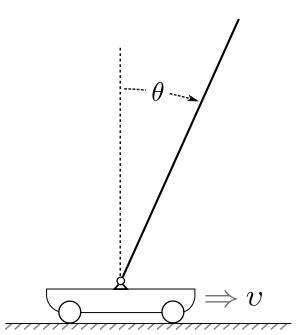}
		\caption{}
		\label{fig:stick} 
	\end{subfigure}
	\begin{subfigure}{0.81\textwidth}
		\centering
		\includegraphics[width=1.0\linewidth]{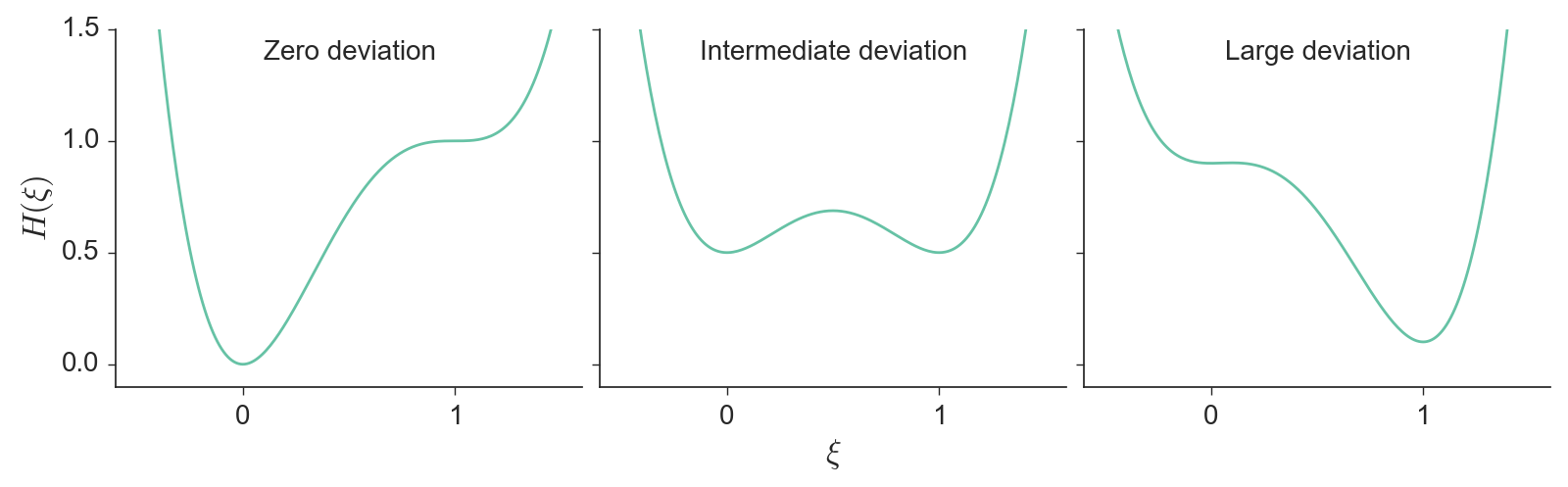}
		\caption{}
		\label{fig:potential}
	\end{subfigure}
	\caption{(a) Overdamped inverted pendulum; (b) Double-well energy landscape $H(\xi)$ depending on deviation of the stick from the desired position.}
\end{figure}

Humans face the task of balancing dynamic systems near an unstable equilibrium repeatedly throughout their lives. The task of inverted pendulum (stick) balancing (Fig.~\ref{fig:stick}) is an increasingly popular paradigm of studying human control behavior in such situations~\cite{balasubramaniam2013control}. In particular, much research has been aimed at understanding the mechanisms of discontinuous, or \textit{intermittent} control in the context of human balance control~\cite{milton2013intermittent, gawthrop2013human}. The present paper deals with one of the recent developments in the theory of human intermittent control, namely, the double-well model of noise-driven control activation.

The analyzed model of inverted pendulum under control of human operator~\cite{zgonnikov2015double} includes three dynamical variables: stick angle $\theta$, cart velocity $\upsilon$, and an order parameter $\xi$ describing the cognitive state of the operator in regards to the controlled system. Their dynamics are defined by the equations
\begin{equation}
\label{eq:model_rescaled}
\begin{aligned}
\dot \theta & = \theta - \upsilon, \\
\dot \upsilon & = \gamma \theta \xi - \sigma \upsilon, \\
\tau \dot \xi & = -\frac{\partial H}{\partial \xi} + \sqrt{\epsilon H} \zeta,  
\end{aligned}
\end{equation}
where $\tau$ is the time scale of the control activation process, $\gamma$ and $\sigma$ are feedback parameters, $H(\xi,\theta)$ is the ``Hamiltonian'' shaping the energy landscape of the control activation process (Fig.~\ref{fig:potential}), $\zeta$ is white noise, and $\epsilon>0$ is the parameter regulating the noise intensity. According to Eqs.~\eqref{eq:model_rescaled}, the order parameter $\xi$ switches intermittently between the states $\xi=0$ and $\xi=1$, which are mapped onto the operator's two cognitive states, ``wait'' and ``act''. In this way, the model captures on-off intermittency observed experimentally in human control.

The double-well model of control activation has previously been shown to describe human behavior in \textit{virtual stick balancing} task, where the operator controls the virtual overdamped stick on a computer screen by observing angular deviation of the stick and moving the computer mouse to adjust the cart position (see Ref.~\cite{zgonnikov2015double} for details of the model and comparison with experimental data). However, recently it has been demonstrated that statistical properties of human behavior in virtual stick balancing can change substantially depending on the type of visual feedback provided to the operator. In particular, when the operator observes not only stick angle, but also linear displacement of the upper tip of the stick from the reference point, the distribution of action points (values of stick angle triggering the operator's response) changes sharply~\cite{zgonnikov2015type}. 

\begin{figure}
	\centering
	\includegraphics[width=0.5\columnwidth]{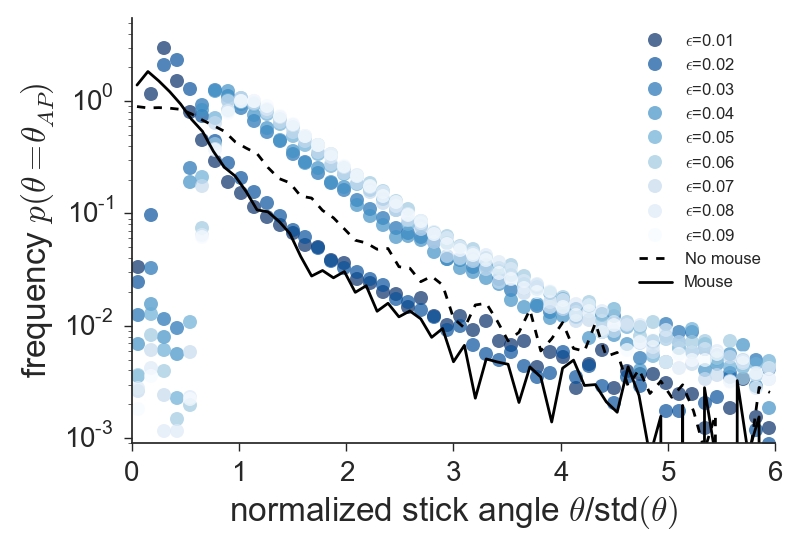}
	\caption{Distributions of action points (AP) exhibited by human operators in virtual stick balancing (black lines) and produced by the double-well model (blue circles). Dashed black line is the AP distribution in angle-only condition (mouse cursor is hidden from the computer screen), and solid black line is the distribution in the linear displacement condition (mouse cursor near the upper tip of the stick on provides additional visual feedback to the operator). For model simulations, only parameter $\epsilon$ was varied, with other parameters taking the values $\sigma=3.5,\,\gamma=\sigma^2/2,\,\tau=0.2$.}
	\label{fig:ap_distr}
\end{figure}

In the present paper we demonstrate that the double-well model can reproduce the whole range of experimentally observed distributions under different conditions. Moreover, we show that a slight change in the model parameter (from $\epsilon=0.02$ to $\epsilon=0.03$) leads to a sudden shift of the action point distribution shape, that is, a phase transition is observed (Fig.~\ref{fig:ap_distr}). 

The obtained simulation results lead us to the hypothesis that the two phases of the system~\eqref{eq:model_rescaled} correspond to two different modes of control activation. The first mode is characterized by Laplace-like action points distribution and high intensity of noise in the control activation mechanism causing repeated switching between on- and off-control periods ($\epsilon \gtrsim 0.03$). This mode is observed experimentally in human operators reacting solely to angular deviations of the stick. The second mode generates power-law-like distributions of action points, and is reproduced by the model for relatively low noise intensity ($\epsilon \lesssim 0.02$). The latter mode is found in the operators supplied with additional sensory feedback. The physiological basis of the two distinct modes needs to be investigated in follow-up studies, which may further shed light on the nature of human intermittent control.

\bibliography{library}
\end{document}